\documentclass[%
 reprint,
 amsmath,amssymb,
 aps,
]{revtex4-2}

\usepackage{graphicx}
\usepackage{dcolumn}
\usepackage{bm}

\begin{document}

\preprint{APS/123-QED}

\title{Generation of four-dimensional hyperentangled N00N states and beyond with photonic orbital angular momentum and detection-basis control}

\author{Jos\'e C\'esar Guerra V\'azquez$^{1,2}$}
 
 \author{Emmanuel Narv\'aez Casta\~neda$^{1,2}$} 
 
 \author{Roberto Ram\'irez Alarc\'on$^{2}$}
 
 \author{Imad Agha$^{1}$}  
 
 \author{Qiwen Zhan$^{3}$}

\author{William N. Plick$^{1}$}%
 \email{wplick1@udayton.edu}
 
\affiliation{%
 $^{1}$Department of Electro-Optics and Photonics, University of Dayton, Dayton Ohio 45469, USA 
}%
\affiliation{%
$^{2}$Centro de Investigaciones en \'Optica A.C., Loma del Bosque 115, Colonia Lomas del Campestre, 37150 Le\'on Guanajuato, M\'exico
}%
\affiliation{%
$^{3}$School of Optical-Electrical and Computer Engineering, University of
Shanghai for Science and Technology, Shanghai 200093, China
}%




\date{\today}

\begin{abstract}
Hyperentanglement of photonic light modes is a valuable resource in quantum information processing and quantum communication. Here we propose a new protocol using the interference of two optical nonlinearities and control of the heralding (detection) basis in the orbital-angular-momentum degree of freedom. This setup is capable of generating states which are both maximally- and hyper- entangled in at least four dimensions. The resultant state in the four-dimensional case is a generalization of the so-called N00N state (a maximally path-entangled state well known in quantum optics). The production of this state is ``perfect'' (in other words noise-less) at least in the ideal case, excluding experimental imperfections. The presented setup is very versatile, and with control of the detection and pumping protocols a massively-large parameter space, of arbitrarily-large dimensionality, may be searched for other states of interest. Also, we present specific cases demonstrating how the state may be tuned from two, to three, to four dimensions - which may be of further theoretical and experimental interest.  
\end{abstract}

\maketitle



\section{Introduction}

Entanglement is perhaps the most-important signature of quantum systems and is essential for the development of quantum technologies, specifically in communications and computing where photonic quantum states are of great interest.
In photonic systems there has been tremendous effort put into studying and implementing quantum protocols such as teleportation and secure cryptography using entangled photons with dimensionalities beyond the two-dimensional limit of polarization, for example: path, photon number, time, frequency, and the complex-spatial degrees of freedom,  \cite{Babazadeh:17,Kues:17, Krenn:2017PhysRevLett.118.080401,Forbes&Nape_2019}.
In the spatial degree of freedom, specifically, there has been a great deal of research focused on the orbital angular momentum (OAM) of entangled photon pairs generated by the spontaneous parametric down-conversion (SPDC) processes \cite{mair2001entanglement, PhysRevA.81.043844, erhard2018experimental}.

In recent years, there have been theoretical and experimental demonstrations of maximally entangled states (MESs) with OAM. The generation of high-dimensional maximally entangled OAM states has been performed using a pump in a superposition of Laguerre-Gauss modes \cite{Kovlakov_2018, Liu_2018}, and later, a broader and flatter OAM spectrum was produced by shaping the pump beam profile \cite{Liu_2020}. 
Applications of high-photon-number N00N states have also been investigated, including angular super-resolution, bunching two photons into different OAM N00N states \cite{hiekkamaki2021photonic}, and quantum information processing using hyperentanglement in many degrees of freedom \cite{deng2017quantum}. Furthermore, a source of hyperentangled states encoded in time-frequency and vector-vortex-structured modes was recently reported \cite{graffitti_2020}, as well as the measurement of a N00N state with $10^{12}$ (a tera) spatio-temporal modes \cite{gao2021teramode}. However, the preparation and measurement of high-dimensional maximally entangled OAM states is still challenging. Especially since the down-conversion process for OAM is inherently ``noisy'' \--- that is \emph{every} possible OAM mode is produced \--- so without post-selection or heralding some unwanted terms in the density matrix are always non-zero.  

In this work, we propose a new method for the generation of hyperentangled OAM N00N states with tunable dimensionality which in the four-dimensional case results in a perfect state. The intrinsic characteristic of this method is interference between two nonlinearities on two beam splitters (Fig. \ref{fig:Seeded-OAM}) with a general heralding (detection) protocol on two of the four resulting modes.

In the following section we review MESs in the spatial degree of freedom. 
In Sec. \ref{sec:Hyperentanglement generation} we present our setup and derive the state for hyperentanglement generation. 
In Sec. \ref{sec:Generating specific OAM N00N states of tunable dimensionality} we present several different specific implementations of the detection-basis control, generating two, three and four-dimensional entangled OAM N00N states. In Sec. \ref{sec:Discussion and future prospects} we discuss future prospects, and in Sec. \ref{sec:Conlusions} we summarize and conclude.

\section{Down-conversion and entangled states with the complex spatial modes of light}

It is well known that as light beams travel through space their phase patterns vectors can rotate about the optic axis. Such light beams carry angular momentum in both the spin angular (SAM) and orbital angular (OAM) degree of freedom. SAM is associated with the polarization and OAM with the azimuthal phase of the complex electric field \cite{pachava2019generation, Zhang_2020}.

Since photon pairs generated by SPDC are entangled in the spatial degree of freedom, they can be described as a superposition of Laguerre-Gauss (LG) modes, which in cylindrical coordinates are described by \cite{Ibarra_2019}

\begin{equation}\label{eq:LG}
\begin{split}
    \mbox{LG}_p^l(\rho, \phi) &= \frac{1}{w_0}
\sqrt{\frac{2 p!}{\pi (p+\vert l \vert)!}} 
\left(\frac{\rho\sqrt{2}}{w_0}\right)^{\vert l \vert} 
L_p^{\vert l \vert}\left(\frac{2\rho^2}{w_0^2}\right)
\\ &\times
\mbox{exp}\left(\frac{-\rho^2}{w_0^2}\right)
\mbox{exp}\left(il\phi\right),
\end{split}
\end{equation}

\noindent where $w_0$ is the beam waist radius at $z=0$, $l$ corresponds to the OAM $l\hbar$ per photon of the beam and describes the helical structure around a wave-front singularity, $p$ is the number of radial nodes in the intensity distribution or, in terms of the intensity cross-section, $p+1$ describes the number of concentric rings of radial intensity maxima, $\phi$ and $\rho$ are the azimuthal and radial coordinates (respectively), and $L_p^{\vert l\vert}(\cdot)$ is the associated Laguerre polynomial.

The entangled state generated by a pump with a single LG-mode 
$\vert l_{\mbox{\tiny{p}}}, p_{\mbox{\tiny{p}}} \rangle$ 
can be written as

\begin{eqnarray}\label{eq:OAM-LG}
\vert\psi_{\mbox{\tiny{SPDC}}}\rangle = 
\sum_{l_s,p_s} \sum_{l_i,p_i} 
B_{p_s,p_i}^{l_s,l_i}\, 
\vert l_s,\,p_s\rangle 
\vert l_i,\,p_i\rangle,
\end{eqnarray}

\noindent where $\vert l_s,\,p_s\rangle$ ($\vert l_i,\,p_i\rangle$) represents a single photon in the signal (idler) mode, the coefficients $\vert B_{p_s,p_i}^{l_s,l_i} \vert^2$ represent the probability to generate a photon pair with signal and idler modes given a pump with LG-mode $\vert l_{\mbox{\tiny{p}}}, p_{\mbox{\tiny{p}}} \rangle$, and the coincidence probability amplitudes of the same are given by the overlap integral \cite{Yao_2011}:

\begin{align}\label{eq:OAM-ampl}
B_{p_s,p_i}^{l_s,l_i} &=
\langle \psi_i, \psi_s \vert \,\psi_{\mbox{\tiny{SPDC}}} \rangle
\nonumber\\
&= \int_0^{2\pi} \mbox{d}\phi \int_0^\infty \rho\,\mbox{d}\rho\,
\mbox{LG}_{p_{\mbox{\tiny{p}}}}^{l_{\mbox{\tiny{p}}}}(\rho,\phi)
[\mbox{LG}_{p_s}^{l_s}(\rho,\phi)]^* 
\nonumber\\
&\times[\mbox{LG}_{p_i}^{l_i}(\rho,\phi)]^*.
\end{align}

The entangled state generated by a pump that is in a superposition of LG-modes is given by the superposition of the states generated by a pump with a single LG-mode, Eq. (\ref{eq:OAM-LG}). So for this case we have

\begin{eqnarray}
\vert \psi_{\mbox{\tiny{pump}}} \rangle = 
\sum_{l_{\mbox{\tiny{p}}},p_{\mbox{\tiny{p}}}} a_{p_{\mbox{\tiny{p}}}}^{l_{\mbox{\tiny{p}}}}\,
\vert l_{\mbox{\tiny{p}}}, p_{\mbox{\tiny{p}}} \rangle.
\end{eqnarray}

Which can be trivially rewritten as

\begin{eqnarray}\label{eq:OAMs-LGs}
\vert\Psi_{\mbox{\tiny{SPDC}}}\rangle = 
\sum_{l_s,p_s} \sum_{l_i,p_i} 
C_{p_s,p_i}^{l_s,l_i}\, 
\vert l_s,\,p_s\rangle 
\vert l_i,\,p_i\rangle ,
\end{eqnarray}

\noindent where

\begin{eqnarray}\label{OAM-ampl-superposition}
C_{p_s,p_i}^{l_s,l_i} = 
\sum_{l_{\mbox{\tiny{p}}},p_{\mbox{\tiny{p}}}} a_{p_{\mbox{\tiny{p}}}}^{l_{\mbox{\tiny{p}}}}\,
B_{p_{\mbox{\tiny{p}}};\,p_s,p_i}^{l_{\mbox{\tiny{p}}};\,l_s,l_i}. 
\end{eqnarray}

\noindent These are the \emph{full} coincidence amplitudes, where $B_{p_{\mbox{\tiny{p}}};\,p_s,p_i}^{l_{\mbox{\tiny{p}}};\,l_s,l_i}$ are the coincidence amplitudes given by Eq. (\ref{eq:OAM-ampl}), calculated for each \emph{individual} Laguerre-Gauss component 
$\vert l_{\mbox{\tiny{p}}}, p_{\mbox{\tiny{p}}} \rangle$
of the initial pump.

\begin{figure}[htbp]
    \centering
    \includegraphics[scale=0.63]{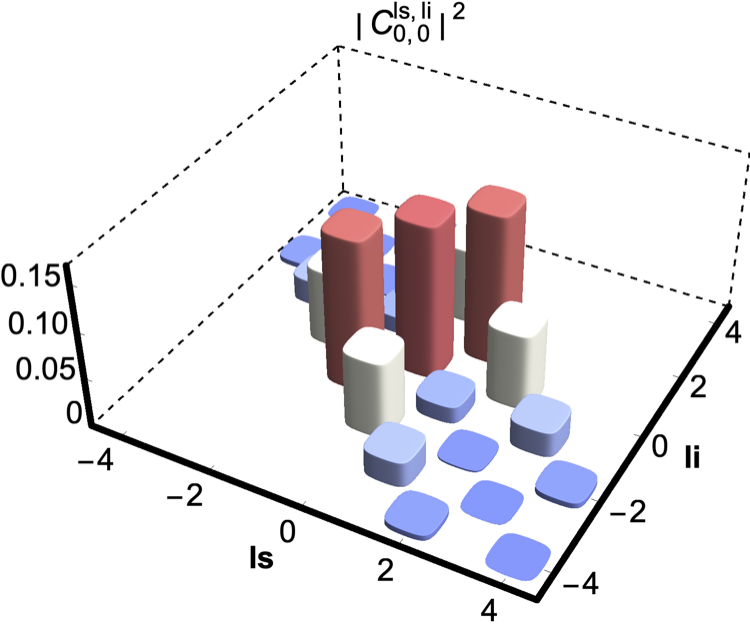}
    \caption{Spiral spectrum distributions of a three-dimensional maximally entangled state (MES). The $x$ and $y$ axes represent the azimuthal values of signal $l_s$ photon and idler $l_i$ photon, respectively, and the $z$ axis represents the coincidence probability amplitudes $\vert C^{l_s, \,l_i}_{0, \,0} \vert^2$ of finding a photon pair with signal and idler modes $\vert l_s,\,l_i\rangle$. Those amplitudes on the diagonal (in red) constitute the target state, whereas other components are unwanted noise.}
    \label{fig:MES3D}
\end{figure}

For example consider the case from \cite{Liu_2018}, where the pump is put into a superposition of three pure azimuthal Laguerre-Gauss modes $\vert l_{\mbox{\tiny{p}}} \rangle$, 
with complex amplitudes $a^{l_{\mbox{\tiny{p}}}}$

\begin{eqnarray}\label{eq:pump-L-modes}
\vert \psi_{\mbox{\tiny{pump}}} \rangle =
N\left(\sqrt{2.5}\,\vert -2 \rangle+\vert 0\rangle+\sqrt{2.5}\,\vert 2 \rangle \right),
\end{eqnarray}

\noindent where $N$ is the normalization constant. The generated entangled state is a three-dimensional maximally entangled state (MES). The resulting probability amplitudes (density matrix) are shown in Fig. \ref{fig:MES3D}. Note that the entangled state here is given by the diagonal elements (shown in red). The other amplitudes constitute unwanted noise. 

\section{Hyperentanglement generation using interference and detection-basis control}\label{sec:Hyperentanglement generation}

\begin{figure}[htbp]
    \centering
    \includegraphics[scale=0.6]{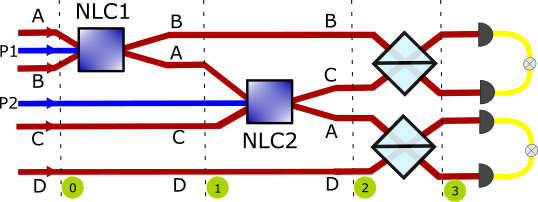}
    \caption{Conceptual diagram of the process. P1 and P2 are the pumps to the non-linearities NLC1 and NLC2, respectively, each producing a down-converted pair of photons into the output modes. 
    Entanglement between photon pairs A and B generated in the first nonlinear crystal (NLC1) is linked to the remaining modes using the second nonlinear crystal (NLC2) in conjunction with the beam-splitters (light-blue boxes) that follow. A coincidence detection protocol using OAM detection-basis control is then used on modes A and D, heralding the desired state between modes B and C. The green icons represent the stages of evolution of the state as it moves through the device.}
    \label{fig:Seeded-OAM}
\end{figure}

In Fig. \ref{fig:Seeded-OAM} we present a diagram of our proposed setup, where P1 and P2 are the pumps to the nonlinearities. The modes A, B, C and D are input modes to the device, originally all in the vacuum state.

The two nonlinear crystals, NLC1 and NLC2, are placed in a non-colinear configuration. We examine both processes in the SPDC regime, i.e. each adds one photon each to the output modes. Entanglement between photon pairs A and B is generated in the first nonlinear crystal (NLC1). The photon in mode A is used as a seed to the second crystal (NLC2). After this point there are assumed to be four photons in the device (this is ensured by post-selection), two in mode A and one each in modes B and C. After that modes A and D, and modes B and C are mixed (interfered) on 50:50 beam splitters. The final step of the procedure is to herald on specific OAM superpositions as described in detail below. The desired state is then produced between modes B and C.

Examining this process in detail we take as our input the vacuum state in modes A, B, C and D. Hence, the initial state of the process (icon zero on Fig. \ref{fig:Seeded-OAM}), before interaction with the nonlinear crystals, is given by

\begin{eqnarray}
\vert \psi_0 \rangle =\vert 0 \rangle  =  \vert 0\rangle_A  \vert 0\rangle_B \vert 0\rangle_C  \vert 0 \rangle_D.
\end{eqnarray}

\noindent After interaction with the first nonlinear crystal (NLC1) the state of the system is described by (icon one on Fig. \ref{fig:Seeded-OAM})

\begin{eqnarray}
\vert \psi_1 \rangle = \sum_l C_{\mbox{\tiny{AB}}}^{-l,l} \,\hat{a}^\dagger_{-l} \hat{b}^\dagger_l \,\vert 0 \rangle,
\end{eqnarray}

\noindent where $C_{\mbox{\tiny{AB}}}^{-l,l}$ are the coincidence amplitudes of the down-converted state, see Eq. (\ref{eq:OAM-ampl}), $\hat{a}^{\dagger}_{-l}$ and $\hat{b}^{\dagger}_{l}$ are the creation operators for the signal and idler modes, respectively, and for convenience we write the negative sign on the signal mode. 

We now use the signal photon produced into the first nonlinear crystal (NCL1) to seed the down-conversion process into the second nonlinear crystal (NLC2). The state of the system after interaction with the second nonlinear crystal is then written as (icon two on Fig. \ref{fig:Seeded-OAM})

\begin{eqnarray}
\vert \psi_2 \rangle = \sum_{\mathcal{L}} \sum_{l} \gamma_{\mbox{\tiny{AC}}}^{\mathcal{-L},\mathcal{L}}\, C_{\mbox{\tiny{AB}}}^{-l,l} \,\hat{a}^\dagger_{-l} \hat{a}^\dagger_{\mathcal{-L}} \hat{b}^\dagger_{l} \hat{c}^\dagger_{\mathcal{L}} \,\vert 0 \rangle,
\end{eqnarray}

\noindent where $\gamma_{\mbox{\tiny{AC}}}^{\mathcal{-L},\mathcal{L}}$ are the coincidence amplitudes of the down-converted state, see Eq. (\ref{eq:OAM-ampl}) and Eq. (\ref{OAM-ampl-superposition}), and $\hat{a}^\dagger_{\mathcal{-L}}$ and $\hat{c}^\dagger_{\mathcal{L}}$ are the creation operators for the signal and idler modes, respectively. 

After beam splitter transformations, and expanding the products we have the state (icon three on Fig. \ref{fig:Seeded-OAM})

\begin{equation}
\begin{split}
    \vert \psi_3 \rangle &= \sum_{\mathcal{L},l} \gamma_{\mbox{\tiny{AC}}}^{\mathcal{-L},\mathcal{L}}\, C_{\mbox{\tiny{AB}}}^{-l,l}\, (\hat{a}^\dagger_{-l} \hat{a}^\dagger_{\mathcal{-L}} -i\hat{a}^\dagger_{-l} \hat{d}^\dagger_{\mathcal{-L}} -i\hat{d}^\dagger_{-l} \hat{a}^\dagger_{\mathcal{-L}} 
    \\
    &-\hat{d}^\dagger_{-l} \hat{d}^\dagger_{\mathcal{-L}})  \times 
    (\hat{b}^\dagger_{l} \hat{c}^\dagger_{\mathcal{L}} -i\hat{b}^\dagger_{l} \hat{b}^\dagger_{\mathcal{L}} -i\hat{c}^\dagger_{l} \hat{c}^\dagger_{\mathcal{L}} -\hat{c}^\dagger_{l} \hat{b}^\dagger_{\mathcal{L}} )
    \,\vert 0 \rangle.
\end{split}
\end{equation}

We can ensure we only have one photon each in modes A and D by heralding only on the coincidence between those two modes, so eliminating terms that do not have such coincidences, we obtain

\begin{equation}
\begin{split}
    \vert \psi_3 \rangle &= \sum_{\mathcal{L},l} \gamma_{\mbox{\tiny{AC}}}^{\mathcal{-L},\mathcal{L}}\, C_{\mbox{\tiny{AB}}}^{-l,l}\, (-i\hat{a}^\dagger_{-l}\hat{d}^\dagger_{\mathcal{-L}} -i\hat{d}^\dagger_{-l} \hat{a}^\dagger_{\mathcal{-L}} )\\ &\times 
    (\hat{b}^\dagger_{l} \hat{c}^\dagger_{\mathcal{L}} -i\hat{b}^\dagger_{l} \hat{b}^\dagger_{\mathcal{L}} -i\hat{c}^\dagger_{l} \hat{c}^\dagger_{\mathcal{L}} -\hat{c}^\dagger_{l} \hat{b}^\dagger_{\mathcal{L}} )
    \,\vert 0 \rangle.
\end{split}
\end{equation}

\noindent Then, for clarity of notation we act these operators onto the vacuum kets ($\vert 0\rangle$), resulting in

\begin{equation}\label{eq:psi3-after-BS}
    \begin{split}
        \vert \psi_3 \rangle = \sum_{\mathcal{L},l} \gamma_{\mbox{\tiny{AC}}}^{\mathcal{-L},\mathcal{L}}\, C_{\mbox{\tiny{AB}}}^{-l,l} (-i\vert-l\rangle_A \vert\mathcal{-L}\rangle_D -i\vert\mathcal{-L}\rangle_A \vert-l\rangle_D)
        \\ \times
        (\vert l\rangle_B \vert\mathcal{L}\rangle_C -i\vert l,\mathcal{L}\rangle_B \vert0\rangle_C -i\vert0\rangle_B \vert l,\mathcal{L}\rangle_C -\vert\mathcal{L}\rangle_B \vert l\rangle_C),
    \end{split}
\end{equation}

\noindent where $\vert l,\mathcal{L}\rangle_B\,(\vert l,\mathcal{L}\rangle_C)$ represents two photons in mode B (C) with OAM $l$ and $\mathcal{L}$.

As we mentioned, the desired state is produced between modes B and C, and the heralding on modes A and D is coincidental using state projectors on specific OAM superpositions.

The key to this protocol is the choice of the projectors used. It is important here to note that it is possible to project on \emph{any superposition} of OAM modes up to an \emph{arbitrarily-large} dimensionality. Any arbitrary superposition of OAM modes is itself a solution to the paraxial wave equation (due to linearity), and thus corresponds to a particular transverse field pattern. If we wish to project on such a mode we can create the inverse of this field pattern (for example with a spatial light modulator \--- SLM), and shine the photon on this pattern. If the incident photon has the desired field pattern then the inverse operation of the SLM will take it (and only it) to the fundamental Gaussian. If the photon is then shone on a single-mode fiber connected to a detector, then since only the fundamental will couple to the fiber, a detector firing indicates a projection onto programmed state. It is important to note that in-principle phase and amplitude modulation is needed, though phase-only modulation could possibly be sufficient. Other experimental implementations allowing projecting on particular states exist as well \cite{Ibarra_2019,Liu_2020,hiekkamaki2021photonic, PhysRevApplied.14.024048}. Mathematically, we use the projector operators on modes A and D given over the general sums of the $j$ and $k$ modes  

\begin{eqnarray}
\label{eq:projectorD}
\vert P\rangle_D = \sum_j f^*_j \vert j\rangle_D,
\\
\label{eq:projectorA}
\vert P\rangle_A = \sum_k g^*_k \vert k\rangle_A,
\end{eqnarray}

\noindent where (ignoring normalization constants) $\vert j\rangle_D=\hat{d}^\dagger_j \vert 0\rangle_D$, $\vert k\rangle_A=\hat{a}^\dagger_k \vert 0\rangle_A$ and, $f^*_j$ and $g^*_k$ are complex amplitudes.
Hence the final state is given by

\begin{equation}\label{eq:psi-f}
\vert \psi_f\rangle = \,_{D}\langle P\vert _{A}\langle P\vert \psi_3\rangle.    
\end{equation}

\noindent Substituting in the projection operators and the state $\vert \psi_3\rangle$ we have

\begin{align}
    \vert \psi_f\rangle &= 
\sum_{\mathcal{L},l} \gamma_{\mbox{\tiny{AC}}}^{\mathcal{-L},\mathcal{L}}\, C_{\mbox{\tiny{AB}}}^{-l,l}\, \sum_{j,k}  f_j \,g_k \, (_{D}\langle j\vert _{A}\langle k\vert) \nonumber\\
&\times (-i\vert-l\rangle_A \vert\mathcal{-L}\rangle_D -i\vert\mathcal{-L}\rangle_A \vert-l\rangle_D)
\\\nonumber &\times
(\vert l\rangle_B \vert\mathcal{L}\rangle_C -i\vert l,\mathcal{L}\rangle_B \vert0\rangle_C -i\vert0\rangle_B \vert l,\mathcal{L}\rangle_C -\vert\mathcal{L}\rangle_B \vert l\rangle_C).
\end{align}

\noindent Then taking the inner product and simplifying we obtain 

\begin{equation}\label{eq:psi-final-AllModes}
    \begin{split}
        \vert \psi_f\rangle &= 
(-i)\sum_{\mathcal{L},l} \gamma_{\mbox{\tiny{AC}}}^{\mathcal{-L},\mathcal{L}} C_{\mbox{\tiny{AB}}}^{-l,l} \sum_{j,k}  f_j g_k 
(\delta_{j,\mathcal{-L}} \delta_{k,-l} 
\\&+\delta_{j,-l} \delta_{k,\mathcal{-L}})
\times
(\vert l\rangle_B \vert\mathcal{L}\rangle_C -i\vert l,\mathcal{L}\rangle_B \vert0\rangle_C 
\\&-i\vert0\rangle_B \vert l,\mathcal{L}\rangle_C -\vert\mathcal{L}\rangle_B \vert l\rangle_C).
    \end{split}
\end{equation}

\noindent The entangled state described by Eq. (\ref{eq:psi-final-AllModes}) is a general state that allows the generation of low-dimensional and high-dimensional entangled states with orbital angular momentum (OAM). Importantly, the dimensionality and structure of the entangled state depends on the number of projector modes we choose and their amplitude values. We describe this process below.

In this expansion we can choose not only how many terms are in each superposition, but also the weights of all of those terms. The state-space produced is \--- in principle \--- infinite and the experimenter is free to chose the size as they wish. This presents both an opportunity and a problem. The opportunity is that it is a near certainty that many more useful states exist in higher dimensions beyond what we present here. The problem is that as dimensionality increases it becomes progressively-harder to search and find these states. This problem is further compounded by the fact that there are many ways to search for \--- and even define \--- entanglement. Also, some states that are entangled in very-many dimensions may be useless for practical applications. Furthermore, the idea of ``closeness'' of quantum states becomes very muddy as dimensionality increases, and a state being close to a useful entangled state does \emph{not} generally mean that that state contains useful entanglement. For further discussion of these points see sections \ref{sec:Generating specific OAM N00N states of tunable dimensionality} and \ref{sec:Discussion and future prospects}.   

\section{Generating specific OAM N00N states of tunable dimensionality}\label{sec:Generating specific OAM N00N states of tunable dimensionality}

In this section we take the result of Sec.  \ref{sec:Hyperentanglement generation} in the form of equations (\ref{eq:projectorD}),  (\ref{eq:projectorA}) and  (\ref{eq:psi-final-AllModes}),
and present several different specific implementations (choices of the detection-basis control) and show that the state can be made perfectly (noiselessly) entangled in two, three, and four dimensions, at minimum. In each of the next three subsections we take progressively more projection modes/dimensions (one in section \ref{subsec:Generation of two-dimensional entangled OAM N00N states}, two in section \ref{subsec:Generation of hyperentangled OAM N00N states}, three in section \ref{subsec:Generation of high-dimensional entangled OAM states}) and show how particular choices of the weights in the mode expansions lead to entanglement in two modes or hyper-entanglement in three or four modes. Note that these values have been chosen ``by inspection'' without numerical optimization (a highly-challenging prospect on its own).

\subsection{Generation of entangled OAM N00N states with single-mode projectors}\label{subsec:Generation of two-dimensional entangled OAM N00N states}

To generate maximally entangled OAM states in two dimensions, we take into account projector operators with a single mode, $j\equiv k=-m$. So that equations (\ref{eq:projectorD}) and (\ref{eq:projectorA}) become

\begin{eqnarray}
\label{eq:projectorD-1mode}
\vert P\rangle_D = f^*_{m} \vert -m\rangle_D,
\\
\label{eq:projectorA-1mode}
\vert P\rangle_A = g^*_{m} \vert -m\rangle_A.
\end{eqnarray}

\noindent Since $j\equiv k=-m$, equation (\ref{eq:psi-final-AllModes}) transforms into 

\begin{align}
        \vert \psi_{f1}\rangle &= 
(-2i)\sum_{\mathcal{L},l} \gamma_{\mbox{\tiny{AC}}}^{\mathcal{-L},\mathcal{L}} C_{\mbox{\tiny{AB}}}^{-l,l}  f_{m} g_{m}
(\delta_{-m,\mathcal{-L}} \delta_{-m,-l})
\nonumber\\\nonumber&\times
(\vert l\rangle_B \vert\mathcal{L}\rangle_C -i\vert l,\mathcal{L}\rangle_B \vert0\rangle_C -i\vert0\rangle_B \vert l,\mathcal{L}\rangle_C 
\\&-\vert\mathcal{L}\rangle_B \vert l\rangle_C),
\end{align}

\noindent which simplifies to

\begin{align}\label{eq:psi-final-1mode}
        \vert \psi_{f1}\rangle &= 
-2 \gamma_{\mbox{\tiny{AC}}}^{-m,m} C_{\mbox{\tiny{AB}}}^{-m,m}  f_{m} g_{m} 
\nonumber\\&\times (\vert m,m\rangle_B \vert0\rangle_C 
+\vert0\rangle_B \vert m,m\rangle_C).
\end{align}

\noindent Clearly, this is an OAM N00N state entangled in a two-dimensional space, which depends on the
weights $f$ and $g$ of the OAM modes that we choose into the state projectors. Though the dimensionality of this state does not beat what can be done \emph{without} the techniques in this paper we present it for completeness, to see how the dimensionality scales with various choices of detection-basis control.

\subsection{Generation of two- or four- dimensional (hyper-)entangled OAM N00N states with two-mode projectors}\label{subsec:Generation of hyperentangled OAM N00N states}

To generate high-dimensional entangled OAM states, we take into account projector operators with two modes, $j\equiv k$. Thus equations (\ref{eq:projectorD}) and (\ref{eq:projectorA}) become

\begin{eqnarray}
\label{eq:projectorD-2modes}
\vert P\rangle_D = f^*_{m} \vert -m\rangle_D +f^*_{n} \vert -n\rangle_D,
\\
\label{eq:projectorA-2modes}
\vert P\rangle_A = g^*_{m} \vert -m\rangle_A +g^*_{n} \vert -n\rangle_A.
\end{eqnarray}

\noindent The general equation described by Eq. (\ref{eq:psi-final-AllModes}) transforms into

\begin{align}\label{eq:psi-final2-TwoModes-deltas}
        \vert \psi_{f2}\rangle &= 
(-i)\sum_{\mathcal{L},l} \gamma_{\mbox{\tiny{AC}}}^{\mathcal{-L},\mathcal{L}} C_{\mbox{\tiny{AB}}}^{-l,l} [ 
2f_{m} g_{m} \delta_{-m,\mathcal{-L}} \delta_{-m,-l} \nonumber\\&+
2f_{n} g_{n} \delta_{-n,\mathcal{-L}} \delta_{-n,-l} \,+\,
(f_{m} g_{n}+f_{n} g_{m})
\nonumber\\&\times 
( \delta_{-m,\mathcal{-L}}  \delta_{-n,-l} +\delta_{-m,-l}\delta_{-n,\mathcal{-L}})
]
\\\nonumber&\times
(\vert l\rangle_B \vert\mathcal{L}\rangle_C -i\vert l,\mathcal{L}\rangle_B \vert0\rangle_C 
\\\nonumber&-i\vert0\rangle_B \vert l,\mathcal{L}\rangle_C -\vert\mathcal{L}\rangle_B \vert l\rangle_C).
\end{align}

\noindent Simplifying, we obtain

\begin{align}\label{eq:psi-final2-TwoModes}
    \vert \psi_{f2}\rangle &= -2[
         \gamma_{\mbox{\tiny{AC}}}^{-m,m} C_{\mbox{\tiny{AB}}}^{-m,m}  f_{m} g_{m} 
         \nonumber\\&\times(\vert m,m\rangle_B \vert0\rangle_C +\vert0\rangle_B \vert m,m\rangle_C) \nonumber\\
        &+ \gamma_{\mbox{\tiny{AC}}}^{-n,n} C_{\mbox{\tiny{AB}}}^{-n,n}  f_{n} g_{n} (\vert n,n\rangle_B \vert0\rangle_C +\vert0\rangle_B \vert n,n\rangle_C) \nonumber\\
        &+ (\gamma_{\mbox{\tiny{AC}}}^{-m,m} C_{\mbox{\tiny{AB}}}^{-n,n}  f_{m} g_{n} +\gamma_{\mbox{\tiny{AC}}}^{-n,n} C_{\mbox{\tiny{AB}}}^{-m,m}  f_{n} g_{m})
        \nonumber\\ &\times 
        (\vert n,m\rangle_B \vert0\rangle_C +\vert0\rangle_B \vert n,m\rangle_C )
        ],
\end{align}

\noindent which is a two-dimensional or four-dimensional entangled OAM state. 
To clarify, consider the case where all the overall amplitude values (which are combinations of $\gamma$s, $C$s, $f$s and $g$s) are set to one. Then the entangled state described by Eq. (\ref{eq:psi-final2-TwoModes}) simplifies to

\begin{equation}\label{eq:psi-final2-TwoModes-Ampl-1}
    \begin{split}
        \vert \psi_{f2} \rangle &= -2[ 
        \vert m,m\rangle_B \vert0\rangle_C +\vert0\rangle_B \vert m,m\rangle_C \\
        &+\vert n,n\rangle_B \vert0\rangle_C +\vert0\rangle_B \vert n,n\rangle_C \\
        &+2(\vert n,m\rangle_B \vert0\rangle_C +\vert0\rangle_B \vert n,m\rangle_C )
        ].
    \end{split}
\end{equation}

\noindent Which we may re-write as

\begin{align}\label{eq:psi-final2-TwoModes-Ampl-1-simplified}
        \vert \psi_{f2} \rangle &= -2[ 
        \vert n,n\rangle_B \vert0\rangle_C 
        +\vert0\rangle_B \vert n,n\rangle_C\nonumber\\
        &+(\vert m\rangle_B +2\vert n\rangle_B) 
        \vert m\rangle_B\vert0\rangle_C 
        \\\nonumber
        &+\vert0\rangle_B \vert m\rangle_C(\vert m\rangle_C +2\vert n\rangle_C)
        ],
\end{align}

\noindent which is again a two-dimensional entangled OAM state.

However, taking the amplitude values $f_m=1,\,g_m=1,\,f_n=1\,\mbox{and}\,g_n=-1$, 
allows us to generate the four-dimensional hyper-entangled OAM N00N state

\begin{equation}\label{eq:psi-final2-TwoModes-4D-N00N}
    \begin{split}
     \vert \psi_{f2} \rangle &= -2[ \gamma_{\mbox{\tiny{AC}}}^{-m,m} C_{\mbox{\tiny{AB}}}^{-m,m}   (\vert m,m\rangle_B \vert0\rangle_C +\vert0\rangle_B \vert m,m\rangle_C) \\
        &- \gamma_{\mbox{\tiny{AC}}}^{-n,n} C_{\mbox{\tiny{AB}}}^{-n,n} (\vert n,n\rangle_B \vert0\rangle_C +\vert0\rangle_B \vert n,n\rangle_C)],
    \end{split}
\end{equation}

\noindent where the OAM modes $m$ and $n$ of two photons are entangled with the vacuum.

So the dimensionality of the entangled state given by Eq. (\ref{eq:psi-final2-TwoModes}) depends on the amplitude values that we choose for $f$ and $g$. With specific choices we can generate  four-dimensional hyperentangled OAM N00N states and  two-dimensional entangled OAM states.

\subsection{Generation of three- or four- dimensional hyper-entangled OAM N00N states with three-mode projectors}\label{subsec:Generation of high-dimensional entangled OAM states}

We now consider the case of projector operators with three modes

\begin{eqnarray}
\label{eq:projectorD-3modes}
\vert P\rangle_D = f^*_{m} \vert -m\rangle_D +f^*_{n} \vert -n\rangle_D +f^*_{r} \vert -r\rangle_D,
\\
\label{eq:projectorA-3modes}
\vert P\rangle_A = g^*_{m} \vert -m\rangle_A +g^*_{n} \vert -n\rangle_A +g^*_{r} \vert -r\rangle_A.
\end{eqnarray}

\noindent So, plugging this in to Eq. (\ref{eq:psi-final-AllModes}) and performing a simple transformation we get

\begin{align}\label{eq:psi-final3-ThreeModes}
    \vert \psi_{f3}\rangle &= -2[
         \gamma_{\mbox{\tiny{AC}}}^{-m,m} C_{\mbox{\tiny{AB}}}^{-m,m}  f_{m} g_{m} 
         \nonumber\\
         &\times(\vert m,m\rangle_B \vert0\rangle_C +\vert0\rangle_B \vert m,m\rangle_C) \nonumber\\
        &+ \gamma_{\mbox{\tiny{AC}}}^{-n,n} C_{\mbox{\tiny{AB}}}^{-n,n}  f_{n} g_{n} (\vert n,n\rangle_B \vert0\rangle_C +\vert0\rangle_B \vert n,n\rangle_C) \nonumber\\
        &+ \gamma_{\mbox{\tiny{AC}}}^{-r,r} C_{\mbox{\tiny{AB}}}^{-r,r}  f_{r} g_{r} (\vert r,r\rangle_B \vert0\rangle_C +\vert0\rangle_B \vert r,r\rangle_C) \nonumber\\
        &+ (\gamma_{\mbox{\tiny{AC}}}^{-m,m} C_{\mbox{\tiny{AB}}}^{-n,n}  f_{m} g_{n} +\gamma_{\mbox{\tiny{AC}}}^{-n,n} C_{\mbox{\tiny{AB}}}^{-m,m}  f_{n} g_{m})
        \nonumber\\ &\times 
        (\vert n,m\rangle_B \vert0\rangle_C +\vert0\rangle_B \vert n,m\rangle_C )
        \\\nonumber
        &+ (\gamma_{\mbox{\tiny{AC}}}^{-m,m} C_{\mbox{\tiny{AB}}}^{-r,r}  f_{m} g_{r} +\gamma_{\mbox{\tiny{AC}}}^{-r,r} C_{\mbox{\tiny{AB}}}^{-m,m}  f_{r} g_{m})
        \\\nonumber &\times 
        (\vert r,m\rangle_B \vert0\rangle_C +\vert0\rangle_B \vert r,m\rangle_C )
        \\\nonumber
        &+ (\gamma_{\mbox{\tiny{AC}}}^{-n,n} C_{\mbox{\tiny{AB}}}^{-r,r}  f_{n} g_{r} +\gamma_{\mbox{\tiny{AC}}}^{-r,r} C_{\mbox{\tiny{AB}}}^{-n,n}  f_{r} g_{n})
        \\\nonumber &\times 
        (\vert r,n\rangle_B \vert0\rangle_C +\vert0\rangle_B \vert r,n\rangle_C )
        ].
\end{align}

\noindent 
The dimensionality of this state depends on the amplitude values that we choose for $f$ and $g$. For example, taking the coincidence amplitudes $\gamma$s and $C$s equal to one, and then choosing the $f$'s and $g$'s equal to one, we obtain

\begin{align}\label{eq:psi-final3-ThreeModes-Ampl-1-simplified}
        \vert \psi_{f3} \rangle &= -2[
        (\vert m\rangle_B +2\vert n\rangle_B) 
        \vert m\rangle_B\vert0\rangle_C \nonumber\\\nonumber
        &+\vert0\rangle_B \vert m\rangle_C(\vert m\rangle_C +2\vert n\rangle_C) \\
        &+(\vert n\rangle_B +2\vert r\rangle_B) 
        \vert n\rangle_B\vert0\rangle_C 
        \\\nonumber
        &+\vert0\rangle_B \vert n\rangle_C(\vert n\rangle_C +2\vert r\rangle_C) \\\nonumber
        &+(\vert r\rangle_B +2\vert m\rangle_B) 
        \vert r\rangle_B\vert0\rangle_C 
        \\\nonumber
        &+\vert0\rangle_B \vert r\rangle_C(\vert r\rangle_C +2\vert m\rangle_C) 
        ],
\end{align}

\noindent which is a three-dimensional entangled OAM state.

If we set $f$'s and $g$'s equal to one with the exception of $g_r=-1$, the state given by Eq. (\ref{eq:psi-final3-ThreeModes}) then becomes

\begin{align}\label{eq:psi-final3-ThreeModes-Ampl-1-simplified-gr}
        \vert \psi_{f3} \rangle &= -2[ 
        \vert n,n\rangle_B \vert0\rangle_C 
        +\vert0\rangle_B \vert n,n\rangle_C \nonumber\\
        &+\vert r,r\rangle_B \vert0\rangle_C +\vert0\rangle_B \vert r,r\rangle_C \\\nonumber
        &+(\vert m\rangle_B +2\vert n\rangle_B) 
        \vert m\rangle_B\vert0\rangle_C 
        \\\nonumber
        &+\vert0\rangle_B \vert m\rangle_C(\vert m\rangle_C +2\vert n\rangle_C)
        ],
\end{align}

\noindent where is evident that the entanglement dimensionality increased to four-dimensions. Thus Eq. (\ref{eq:psi-final3-ThreeModes-Ampl-1-simplified-gr}) represents a four-dimensional entangled OAM state.

Similarly, taking $g_n=-1$ or $g_m=-1$ allows us to generate the following four-dimensional entangled OAM states

\begin{align}\label{eq:psi-final3-ThreeModes-Ampl-1-simplified-gn}
        \vert \psi_{f3} \rangle &= -2[ 
        \vert m,m\rangle_B \vert0\rangle_C 
        +\vert0\rangle_B \vert m,m\rangle_C \nonumber\\
        &+\vert n,n\rangle_B \vert0\rangle_C 
        +\vert0\rangle_B \vert n,n\rangle_C \\\nonumber
        &+(\vert r\rangle_B +2\vert m\rangle_B) 
        \vert r\rangle_B\vert0\rangle_C 
        \\\nonumber
        &+\vert0\rangle_B \vert r\rangle_C(\vert r\rangle_C +2\vert m\rangle_C)
        ],
\end{align}

\noindent and

\begin{align}\label{eq:psi-final3-ThreeModes-Ampl-1-simplified-gm}
        \vert \psi_{f3} \rangle &= -2[ 
        \vert m,m\rangle_B \vert0\rangle_C 
        +\vert0\rangle_B \vert m,m\rangle_C \nonumber\\
        &+\vert r,r\rangle_B \vert0\rangle_C +\vert0\rangle_B \vert r,r\rangle_C \\\nonumber
        &+(\vert n\rangle_B +2\vert r\rangle_B) 
        \vert n\rangle_B\vert0\rangle_C 
        \\\nonumber
        &+\vert0\rangle_B \vert n\rangle_C(\vert n\rangle_C +2\vert r\rangle_C)
        ].
\end{align}

\noindent Again, the entanglement dimensionality of the general state $\vert \psi_{f3} \rangle$ (Eq. (\ref{eq:psi-final3-ThreeModes})) depends on the 
weights that we choose for each mode on the state projectors 
$\vert P\rangle_D$ and $\vert P\rangle_A$ given by Eq. (\ref{eq:projectorD-3modes}) and Eq. (\ref{eq:projectorA-3modes}), respectively.

\section{Discussion and future prospects}\label{sec:Discussion and future prospects}

Throughout this manuscript we have several times made the distinction between entanglement that is ``useful'', and entanglement that is not. Briefly, we call a state useful if it can directly be applied to some protocol in quantum communication, computation, or metrology \--- for example: mutually-unbiased-basis states, Bell states, N00N states, Greenberger–Horne–Zeilinger states, etc. Entanglement in many-more dimensions than we present here has been found in the spatial degree of freedom of SPDC. For example in \cite{krenn_huber_fickler_lapkiewicz_ramelow_zeilinger_2014} entanglement was found in 100 by 100 dimensions. However in such cases entanglement is found by performing very, very many measurements which must be searched through with a difficult procedure to uncover the entanglement. Though it can be shown that entanglement in such high-dimensionality does indeed exist naturally in SPDC, there does not exist any clear way (to the best of our knowledge) to see how such entanglement could be utilized for a practical technological outcome. In such situations it is difficult to say what the underlying state that contains the entanglement found even is (in contrast to what is noise). They are far from maximally-entangled. N00N states, however, are one of the most sought-after possibilities.

Having devised a new protocol for the (theoretically) noise-less generation of high-dimensional hyperentangled OAM N00N states with detection-basis control, a number of new research avenues become accessible. However, the current limitation to expanding beyond the four-dimensional case to higher dimensions is the optimization process needed.

In our method of detection-basis control the state in both heralding modes can be projected onto any superposition of OAM modes with any dimensionality, i.e., we can choose both the number of modes in the superposition and their weights, which is the main characteristic of our protocol. However, in the three cases we presented here all the choices were made by inspection, hence an immediate next step is to go to higher dimensions (more terms) and optimize the weights of the projectors used to create other states. 

This is no easy task, unfortunately. First there is the massive parameter space that is available to search. Though this is an opportunity, the problem of optimizing in high dimensions is well known. Not only is there the trouble of running a minimization/maximization procedure with many parameters, but also in higher dimensions most concepts of ``distance'' converge on similar values (for example, since the projection of one vector onto another is essentially an averaging procedure over each dimension, as dimensionality becomes large these ``average'' values tend towards being the same).

We, so far, have presented only states that are theoretically ``perfect'' and contain no unwanted terms. If we wish to generalize to more dimensions it's nearly-certain such undesirable terms will show up. However some terms will degrade the entanglement (those that are cross-terms between the dimensions of the target state allowing the state to partially factorize), and some are merely noise. A direct state projection, for example, would not distinguish between these two cases. So secondly, there is the problem of finding a well-formed and useful metric for whatever optimization process is used.

Whether or not a state that is ``close'' to an entangled state itself has useful entanglement is a complicated question, and only recently has appreciable progress been made. If a state does contain such entanglement it is called ``faithful''. Papers have come out recently studying the faithfulness of entanglement \cite{PhysRevLett.126.140503, hu2020optimized}. However deciding the faithfullness of an entangled state is \emph{itself} an optimization procedure, so any search will involve a double optimization. 

It may also be beneficial to study the effects of controlling the pump beams as well, offering full control of the generated state \--- or to make the state ``even more hyper'' by exploiting other degrees of freedom such as polarization or frequency. It is also likely that machine learning techniques, which excel at optimization and pattern recognition, could be applied to the search for other useful entangled states.

Research in these directions is underway.

\section{Conclusions}\label{sec:Conlusions}

In this work we present a new protocol for the generation of hyperentangled OAM N00N states with tunable dimensionality using a simple technique we call detection-basis control. 

In our setup we use the interference of two optical nonlinearities and a coincidence detection protocol at the output of two beam splitters. In this configuration, we show that the down-converted OAM state is perfectly entangled in two, three and four dimensions, at minimum.
We also show how we can tune the dimensionality by heralding on specific OAM superpositions.
Hence, the protocol creates new access to high-dimensional hyperentangled OAM N00N states up to four dimensions, and potentially many more.

Our protocol is a potential resource of high-dimensional hyperentangled OAM N00N states for applications in quantum technologies, such as metrology, sensing, imaging, computing or communication where high-dimensional N00N states are desirable.

\section*{Funding}
This work was supported by CONACYT, Mexico (grants 293471,  293694, Fronteras de la Ciencia 217559).

\section*{Acknowledgments}
We acknowledge support from CONACYT, Mexico.

\nocite{*}

\bibliography{apssamp}

\end{document}